\documentclass[11pt]{article}

\usepackage{graphicx}
\usepackage[utf8]{inputenc}
\usepackage[dvipsnames]{xcolor}
\usepackage{nicefrac}
\usepackage{amsthm}
\usepackage{comment}
\usepackage{subfigure}
\usepackage{mathtools}
\usepackage[style=apa,bibstyle=apa,doi=false,url=false,isbn=false,eprint=false,sortcites=false]{biblatex}
\usepackage[colorlinks=true, urlcolor=blue, linkcolor=blue, citecolor=red]{hyperref}
\usepackage[noabbrev,nameinlink,capitalise]{cleveref}
\usepackage{multirow,array}
\usepackage[margin=1in]{geometry}
\usepackage{setspace}
\usepackage{url}
\usepackage{authblk}
\usepackage{paralist}
\usepackage{bm}
\usepackage{bbm}
\usepackage{amsmath,amssymb}
\usepackage{tikz}
\usepackage{dirtytalk}

\newtheorem{proposition}{Proposition}[section]

\newtheorem{claim}{Claim}[section]
\newtheorem{cor}{Corollary}[section]
\DeclareMathOperator*{\argmax}{arg\,max} 	
\newcommand{\ind}[1]{\mathbf{I}_{\{#1\}}}

\allowdisplaybreaks

\theoremstyle{definition}
\newtheorem{definition}{Definition}[section]
\newtheorem{example}{Example}[section]

\title{Matching Under Preference Uncertainty:\\
Random Allocation, Informativeness, and Welfare}

\author[1]{Yu-Ting Ho\thanks{\href{mailto:hoyuting@berkeley.edu}{hoyuting@berkeley.edu}}}
\affil[1]{Department of Economics, University of California, Berkeley}

\date{\today}

\begin{document}

    \maketitle

    \begin{abstract}

This paper studies a decentralized many-to-one matching market where preferences remain uncertain during the matching process. Institutions initiate matching by sending offers, and applicants decide whether to accept upon receiving them. Since applicants learn their preferences only after receiving offers, institutions struggle with deciding how many offers to issue. I address this challenge by introducing probabilistic offers—admitting applicants with a probability less than one—which ensure that ex-ante market clearing and stability are achievable. However, the welfare effect of information is subtle: applicants may become worse off as they acquire more information.
        
    \end{abstract}
     \newpage
    \section{Introduction}
    Most many-to-one matching markets operate in a decentralized manner, such as graduate admissions and hiring processes in the private sector. In these markets, although applicants initially apply to institutions, the matching is essentially proposed from the institutions’ side. Institutions send out offers to applicants according to their priority ranking and capacity, and thus an applicant may hold multiple offers simultaneously. The final matching is formed only after applicants choose which offer to accept.

A common feature of such markets is that applicants have uncertainty about their own preferences and often learn about them only after receiving offers. A motivating example is the graduate admission process: students typically begin to learn more about programs and their preferences during visit days, which occur after admissions have been issued. Given that there is a capacity constraint for institutions, they face the challenge of determining how many offers to send out, as applicants may update their beliefs about their preferences in the period between receiving offers and making their final decisions. 

This paper proposes a potential solution to this problem by introducing random allocation of offers. In my framework, institutions are allowed to issue probabilistic offers to the last applicants they wish to include—an idea that corresponds to waitlists in practice. I show that, by choosing appropriate cutoffs (i.e., determining how many offers to send out according to priority), the expected number of accepted offers for each institution equals its capacity.

The second part of the paper examines how the information structure affects applicants’ welfare. It is natural to expect that applicants’ welfare improves as they receive more informative signals about their preferences. However, the effect of information on welfare is actually subtle. Even if all applicants obtain weakly more information, it is possible for everyone to become worse off. Moreover, an applicant who receives strictly more information may strictly decrease her own welfare. 

Matching markets with complete information that operate in a decentralized manner have been shown to achieve stable outcomes characterized by cutoff structures (\textcite{kelso1982job}; \textcite{azevedo2016supply}). However, in reality, uncertainty about applicants’ preferences is inevitable, and uncertainty regarding offer acceptance poses significant challenges for institutions. To address this issue, institutions often adopt policies such as early admissions or short decision windows to secure commitments from applicants more efficiently. Yet, as noted by \textcite{avery2007new}, these policies can create additional instability and chaos in the matching process, leading to worse outcomes.

Some studies have attempted to address this problem, but most have arrived only at negative results. The main difficulty lies in defining an appropriate notion of stability. In the canonical setting, stability consists of two properties: respecting capacity and the absence of blocking pairs. However, once preference uncertainty is introduced into the matching model, these two properties are no longer well-defined. 

This paper circumvents the difficulty by introducing an ex-ante concept. There are two layers of uncertainty in my model. First, instead of sending deterministic offers, institutions may issue probabilistic offers, which introduce uncertainty in each applicant’s choice set. Second, conditional on a given choice set, each applicant faces uncertainty regarding their own preferences. Under different information structures, applicants may choose different institutions as signals and states are realized, due to updates in their posterior beliefs. Consequently, a matching can be represented as a randomization over deterministic outcomes, depending on which deterministic cutoffs and signals are realized.

Under the cutoff structure, rather than participating in a centralized mechanism, each applicant simply makes choices under uncertainty, and the decisions of other applicants do not affect her own outcome. Note that institutions must issue offers sequentially according to their priority order: if an applicant does not hold an offer from a given program, no lower-priority applicant can receive an offer from that program. Given that applicants always select the institution that maximizes their expected utility for each realized posterior belief, for any matching, if an applicant wishes to be allocated to a program with a higher probability, that program will not extend an offer to any applicant with lower priority. This property corresponds to the no-blocking condition in the standard definition of stability and is inherently satisfied within my framework. The remaining element required for stability is the capacity constraint.

By applying the Knaster–Tarski theorem, I show that for each information structure, there exists an ex-ante cutoff—defined as the expected value of the deterministic cutoffs that each program may apply—that clears the market ex-ante. In other words, the aggregate expected demand for each program, accounting for the probabilities of different states and signals being realized, equals its capacity. This result corresponds to the capacity constraint in the traditional notion of stability. Allowing probabilistic offers is essential for achieving market clearing; if programs are restricted to deterministic cutoffs, the market may fail to clear.

Under the ex-ante notion of stability, the canonical results from the literature continue to hold. These include the Rural Hospital Theorem and the lattice structure of market-clearing cutoffs, which correspond to students’ welfare. The next question of interest is how the information structure shapes students’ welfare.

In the subsequent analysis, I treat the information structure as exogenously given and establish the corresponding results, since the way individuals learn about their preferences cannot be directly designed. Nevertheless, welfare can be compared across matchings induced by different information structures. It may be conjectured that students’ welfare should increase in the Pareto order as the information structure becomes more informative. This, however, is not generally true, since the Blackwell theorem does not apply in this setting. The relationship between information and welfare is inherently subtle and depends on the specific interaction between market-clearing cutoffs and different beliefs that may be induced.

As applicants receive different signals, their posterior beliefs and induced choices vary across possible choice sets, thereby influencing the market-clearing cutoffs. Yet, these cutoffs do not change monotonically with the informativeness of the information structure. Hence, although applicants may form more accurate beliefs about their preferences under a more informative signal, they might face smaller choice sets due to lower market-clearing cutoffs, which can eventually reduce their expected welfare. The Blackwell theorem applies only to fixed choice problems under uncertainty, whereas in my model, the interaction between signal-induced decisions and the endogenously determined choice sets breaks this separability, rendering the theorem inapplicable. 

Through an example, I demonstrate the possibility that while all applicants receive weakly more informative signals, they may all end up with weakly worse outcomes. In particular, an applicant who obtains strictly more information may receive a strictly lower expected utility. Moreover, even when signals reveal complete information to all applicants, it does not necessarily lead to a Pareto efficient outcome. To ensure that a fully revealing signal induces a Pareto efficient outcome, it is necessary to assume that all institutions share the same priority. These results highlight the complexity of how information affects welfare in matching markets, especially when institutions have heterogeneous priorities.

In summary, this paper provides a potential solution for matching markets in which applicants acquire information about their preferences after the cutoff is released. The main result applies to any information structure, accommodating various ways through which individuals learn about their preferences. The necessity of probabilistic offers provides a theoretical explanation for the widely implemented waitlist policy in real-world matching markets. Although I do not obtain a positive result regarding the relationship between information structures and students’ welfare, my analysis reveals an unexpected possibility—that more information may not always lead to better outcomes. The interaction between signals and market-clearing cutoffs in my framework offers insight into how information may, under certain conditions, reduce students’ welfare.

\subsection{Related Literature}

Matching theory has been widely applied to problems such as school choice, college admissions, and decentralized job matching markets since several seminal works (\textcite{abdulkadirouglu2003school}, \textcite{abdulkadirouglu2009strategy}, \textcite{gale1962college}, and \textcite{kelso1982job}). A growing literature delves into matching markets with uncertain preferences. Broadly, there are two strands of research in this area. The first considers settings in which each individual perfectly knows their own preferences, while the other side of the market has only partial information about them. The second strand, which this paper focuses on, studies environments where even individuals themselves face uncertainty about their own preferences.

In the first strand, a common approach is to apply game-theoretic methods. \textcite{che2016decentralized} examines a college admissions problem in which schools have incomplete information about students’ preferences and incur negative payoffs when enrolling students beyond capacity. They allow schools to admit students without the restriction that offers must be sent sequentially according to priority order and derive equilibrium strategies, although their result is limited to the case with two schools. In contrast, my paper does not model the strategic behavior of schools. Instead, I derive the ex-ante stable matching under different information structures. \textcite{ehlers2015matching} derive the necessary and sufficient conditions for a strategy profile to constitute an ordinal Bayesian Nash equilibrium under incomplete information in a stable mechanism.

In the second strand of research, the notion of stability under incomplete information plays a central role and has been examined in several studies (\textcite{chakraborty2010two}, \textcite{liu2014stable}, \textcite{bikhchandani2017stability}, \textcite{liu2020stability}, \textcite{chen2020learning}, and \textcite{kloosterman2020school}). Related work in computer science explores similar problems using different approaches and methodologies. For example, \textcite{aziz2016stable} investigates how to maximize the probability of achieving stability under the traditional notion, while \textcite{dai2021learning} and \textcite{liu2015two} study agents’ decision-making problems under preference uncertainty.

Some studies incorporate information acquisition (\textcite{kamenica2011bayesian}; \textcite{gentzkow2016rothschild}) into this second strand of research. \textcite{bade2015serial} introduces endogenous information acquisition into the house allocation problem and shows that serial dictatorship is the unique ex-ante Pareto-optimal, strategy-proof, and non-bossy mechanism. \textcite{immorlica2020information} propose a new notion of stability that accounts for costly information acquisition and prove the existence of stable matchings under this framework. \textcite{escobar2024search} examines centralized school assignment mechanisms and uses a game-theoretic approach to study how information acquisition affects equilibrium behavior and welfare.

Last but not least, an increasing number of empirical studies examine the impact of preference uncertainty and information acquisition in matching markets and analyze the associated welfare implications (\textcite{arteaga2022smart}, \textcite{grenet2022preference}, \textcite{hastings2008information}, \textcite{kapor2020heterogeneous}). These works provide empirical evidence that highlights the importance of understanding matching under uncertainty.

The remainder of the paper is organized as follows. Section 2 introduces the model, the setup of uncertainty, and the cutoff structure. Section 3 illustrates the matching process and its outcomes under my framework, concluding with an example. Section 4 presents the main results of the paper. Section 5 discusses the effect of the information structure on students’ welfare. Section 6 concludes.

    \section{Model}
    There is a finite set 
$S = \{s_1, \dots, s_t\}$ of students and a finite set $P = \{p_1, \dots, p_n\}$ of programs. Each student $s_k$ holds a strict preference $\succ_k$ over the set of programs $P$. Let $\succ_S =  (\succ_k)_{1 \le k \le t}$ denote the preference profile of students. Each program $p$ holds a strict priority $\mathcal{R}_p$ over the set of students $S$. Each program $p$ is associated with a capacity constraint $M_p$. $M_p$ represents the  number of students that $p$ aims to acquire.

\subsection{Preference Uncertainty and Information Structure}

 I assume that students do not know their exact preferences, but share a common prior over the set of possible preference profiles. Let $\Omega$ be the set of states of the world. Each $\omega \in \Omega$ is associated with a preference profile $\succ_S$ of the students. I use $\psi \in \Delta(\Omega)$ to denote the common prior over $\Omega$. 

Each student $s_k$ is endowed with a  signal $\pi_k$ to acquire information about her preferences. I assume that signals are exogenously given, since we cannot design how people learn about their preferences. A signal $\pi_k$ consists of a finite realization space $I_k$ and a family of distributions $\{ \pi_k(\cdot|\omega) \}_{\omega \in \Omega}$ over $I_k$. Each signal realization $i_k \in I_k$ induces a posterior belief $\mu^{i}_k \in \Delta(\Omega)$. Accordingly, a signal induces a distribution over posterior beliefs, denoted by $\tau_k \in \Delta(\Delta(\Omega))$. A signal $\pi_k$ induces $\tau_k$ if $\operatorname{Supp}(\tau_k) = \{ \mu^i_k \}_{i_k \in I_k}$ and
\begin{align*}
    \mu^{i}_k (\omega) &= \frac{\pi_k(i_k|\omega) \psi(\omega)}{\sum_{\omega' \in \Omega}\pi_k(i_k|\omega') \psi(\omega')} \quad \text{for all } i_k \in I_k, \ \omega \in \Omega, \\
    \tau_k(\mu_k) &= \sum_{i_k: \mu^{i}_k = \mu_k} \sum_{\omega' \in \Omega}\pi(i_k|\omega') \psi(\omega') \quad \text{for all } \mu_k.
\end{align*}
We say that a belief $\mu_k$ is induced by a signal $\pi_k$ if $\tau_k$ is induced by that signal $\pi_k$ and $\tau_k(\mu_k) > 0$. A distribution of posteriors is Bayes plausible if the expected posterior equals the prior:
\[
\sum_{\mu_k \in \operatorname{Supp}(\tau_k)} \mu_k \, \tau_k(\mu_k) = \psi.
\]

Let $\pi = (\pi_k)_{1 \le k \le t}$ denote the signal profile. Signals are realized after admissions are issued (i.e., after the cutoff is released), and students then choose among the programs that admit them based on their posterior beliefs. The next subsection introduces how programs issue admissions by releasing cutoffs.

\subsection{Cutoff and Choice Set}

Based on the priority relation $\mathcal{R}_p$ of each program $p$, we can establish the ranking of each student $s_k$ with respect to $p$. Let $
r(s_k) = (r_1(s_k), \dots, r_n(s_k)) \in \mathbb{N}^n$ 
be the mapping that specifies the rank of $s_k$ under each program $p$, where  
\[
r_p(s_k) = \bigl|\{s \in S : s \ \mathcal{R}_p \ s_k\}\bigr| \quad \forall \, p \in P.
\]

Admission will be issued sequentially according to students’ rankings. A deterministic cutoff $\bar{b} = (\bar{b}_1, \dots, \bar{b}_n) \in \mathbb{N}^n$ specifies how many students each program will admit. Formally, $s_k$ is admitted to $p$ under the deterministic cutoff $\bar{b}$ if $r_p(s_k) \le \bar{b}_p$. Let $\bar{B}$ denote the set of deterministic cutoffs. Once the deterministic cutoff is announced, each student $s_k$ will have a set of programs that admit them. Let $C : S \times \bar{B} \to 2^P$ denote the choice set function, where  
\[
C(s_k, \bar{b}) = \{p \in P : r_p(s_k) \le \bar{b}_p\}
\]  
is the choice set of student $s_k$ under the deterministic cutoff $\bar{b}$.

I allow a program to issue a probabilistic offer to the last student it wishes to admit. That is, the student will be admitted to the program with some probability strictly between $0$ and $1$. Let $s_k$ be the last student that program $p$ intends to admit. Program $p$ can adopt a randomization over deterministic cutoffs, denoted by $\phi_p \in \Delta\{1, \dots, t\}$, where $\phi_p$ is restricted by the constraint
\[
\operatorname{Supp}(\phi_p) = \{r_p(s_k)-1, \, r_p(s_k)\}.
\]
Let $b_p$ denote the expected number of students that program $p$ admits according to $\phi_p$.

An \emph{ex-ante cutoff} $b = (b_1, \dots, b_n) \in \mathbb{R}^n_{\ge 0}$ specifies the expected number of students each program $p$ will admit. Given that each program can issue a probabilistic offer for its last admitted student, we can associate each ex-ante cutoff $b$ with a randomization $\phi^b_p \in \Delta\{1, \dots, n\}$ adopted by each program $p$, where
\[
\phi^b_p(\bar{b}_p) =
\begin{cases}
    b_p - (\bar{b}_p - 1), & \text{if } 0 \le \bar{b}_p - b_p \le 1, \\[6pt]
    1 - (b_p - \bar{b}_p), & \text{if } -1 \le \bar{b}_p - b_p < 0, \\[6pt]
    0, & \text{otherwise}.
\end{cases}
\]

For illustration, let $b_p = 1.7$. This implies that program $p$ will adopt cutoff $1$ with probability $0.3$ and cutoff $2$ with probability $0.7$.  

We can further define a distribution over the set of deterministic cutoffs, $\phi^b \in \Delta\bar{B}$, associated with the ex-ante cutoff $b$, where
\[
\phi^b(\bar{b}) = \prod_{p \in P} \phi^b_p(\bar{b}_p), \quad \forall \, \bar{b} \in \bar{B}.
\]

In this setting, I assume that programs randomize their deterministic cutoffs independently, and each ex-ante cutoff $b$ corresponds to a unique distribution $\phi^b$. Under the ex-ante cutoff $b$, the probability that student $s_k$ receives the choice set $C(s_k, \bar{b})$ is given by $\phi^b(\bar{b})$. Thus, the ex-ante cutoff serves as a random allocation of admission for students.  

One real-world application of this mechanism is the use of waitlists. Instead of rejecting students outright, programs can place them on a waitlist, under which they are admitted with some probability.

    \section{Matching with Random Allocation}
    There are two key elements that distinguish my paper from the existing literature. First, I incorporate preference uncertainty and an information structure to capture real-world problems. Second, I introduce random allocation of admissions, which serves as a tool to address preference uncertainty in matching markets. Before presenting the main results, this section connects the concepts of information structure and random allocation with the existing matching literature. I begin by describing the matching process, then outline the students’ decision rules, and conclude with an illustrative example.

\subsection{Matching Process}

In the first step, each program $p$ learns its priority $\mathcal{R}_p$, while each student $s_k$ learns the common prior $\psi$ over the set of possible preference profiles and receives a signal $\pi_k$. Next, programs announce an ex-ante cutoff $b$, which is associated with the distribution $\phi^b$, determined by the prior $\psi$ and the signal profile $\pi$. The cutoff distribution $\phi^b$ is realized, and each student receives a signal realization $i_k$ from the signal profile $\pi$. Let $\bar{b}$ denote the realized deterministic cutoff, and let $i_k$ be the signal realization received by each student $s_k$. Each student $s_k$ makes a decision based on the realized choice set $C(s_k,\bar{b})$ and the posterior belief $\mu^{i_k}$. Once students make their decisions, the deterministic matching is formed.

\subsection{Decision Rule}

After observing the realizations of the ex-ante cutoff $b$ and the signal profile $\pi$, students still face uncertainty about their preferences. To evaluate the expected payoff from admission to each program, I adopt a cardinal utility framework. The utility of student $s_k$ is given by $
u_k: P \times \Omega \to \mathbb{R}$,
where $u_k(p,\omega)$ denotes the payoff of attending program $p$ under state $\omega$. For each $s_k$, it holds that $u_k(p,\omega) > u_k(p',\omega)$ if and only if $p \succ_k p'$ under $\omega$, for all $p, p' \in P$ and $\omega \in \Omega$.

The decision rule of student $s_k$ is defined as a function $
\sigma_k: 2^P \times I_k \to P$, 
where $\sigma_k(C, i_k)$ denotes the program chosen when her choice set is $C$ and her signal realization is $i_k$. Each student is assumed to select the program in her choice set that maximizes expected payoff under her posterior belief, i.e.,
\[
\sigma_k(C,i_k) = \argmax_{p\in C} \sum_{\omega \in \Omega} \mu^{i_k}_k(\omega)\, u_k(p,\omega),
\]
where $\mu^i_k$ is the posterior induced by signal $i_k$. I assume that no two programs yield the same expected payoff for a student under any posterior, or under the prior, which ensures strict preferences in terms of expected payoff.

The decision rule defined above is deterministic, representing a special case of decision rules satisfying the obedience condition in \textcite{bergemann2016bayes}. In the more general setting, the decision rule may instead be probabilistic, assigning distributions over the choice set. My main results extend to this more general case, but for clarity, I present the results under the simplifying assumption that students behave as naive utility maximizers. A detailed discussion and the corresponding results for the general case are provided in the appendix.

\subsection{An Example}

After introducing the decision rule, each pair of ex-ante cutoff $b$ and signal profile $\pi$ induces a distribution over deterministic matchings. I illustrate this idea with the following example.  

Let $P = \{p_1,p_2\}$ and $S = \{s_1,s_2\}$, and suppose both programs share the same priority ordering $s_1 \mathcal{R}_p s_2$. Given the common prior $\psi$, the preference profiles of students and their associated states are:  
\begin{align*}
    \omega_1: \ & s_1: p_1 \succ p_2, \quad s_2: p_1 \succ p_2, \\
    \omega_2: \ & s_1: p_2 \succ p_1, \quad s_2: p_1 \succ p_2.
\end{align*}

Consider the signal profile $\pi$ that fully reveals states for both students, and an ex-ante cutoff $b$ composed of $\bar{b}^1 = (1,2)$ and $\bar{b}^2 = (2,2)$. We can write $\pi_k(i_1|\omega_1) = \pi_k(i_2|\omega_2) = 1$ for $k=1,2$, and $
b = \phi^b(\bar{b}^1)\bar{b}^1 + \phi^b(\bar{b}^2)\bar{b}^2.$

When $\bar{b}^1$ is realized, $s_1$ has the choice set $C(s_1,\bar{b}^1) = P$, while $s_2$ has the choice set $C(s_2,\bar{b}^1) = \{p_2\}$. If the state is $\omega_1$ and $i_1$ is realized, then $s_1$ chooses $p_1$ to maximize her utility, and $s_2$ is assigned to $p_2$ since her choice set is a singleton. On the other hand, if the state is $\omega_2$ and $i_2$ is realized, then $s_1$ chooses $p_2$, while $s_2$ remains assigned to $p_2$. A similar derivation applies when $\bar{b}^2$ is realized.  

The probability that each student is matched with each program can be represented in the following matrix:
\[
\bordermatrix{
    & p_1 & p_2 \cr
s_1 & \phi^b(\bar{b}^1)\psi(\omega_1) + \phi^b(\bar{b}^2)\psi(\omega_1) 
    & \phi^b(\bar{b}^1)\psi(\omega_2) + \phi^b(\bar{b}^2)\psi(\omega_2) \cr
s_2 & \phi^b(\bar{b}^2)\psi(\omega_1) + \phi^b(\bar{b}^2)\psi(\omega_2) 
    & \phi^b(\bar{b}^1)\psi(\omega_1) + \phi^b(\bar{b}^1)\psi(\omega_2) \cr
}.
\]

Here, rows correspond to students and columns to programs. Each element gives the probability that the student in the row is matched to the program in the column. Equivalently, this matrix can be expressed as the weighted sum of deterministic matchings:  
\[
\phi^b(\bar{b}^1)\psi(\omega_1)\begin{pmatrix}
1 & 0 \\ 0 & 1
\end{pmatrix}
+
\phi^b(\bar{b}^1)\psi(\omega_2)\begin{pmatrix}
0 & 1 \\ 0 & 1
\end{pmatrix}
+
\phi^b(\bar{b}^2)\psi(\omega_1)\begin{pmatrix}
1 & 0 \\ 1 & 0
\end{pmatrix}
+
\phi^b(\bar{b}^2)\psi(\omega_2)\begin{pmatrix}
0 & 1 \\ 1 & 0
\end{pmatrix}.
\]

Each term above represents the realized probability of a pair of deterministic cutoffs and signal realizations together with its corresponding deterministic matching. For example, the deterministic matching in which $s_1$ attends $p_1$ and $s_2$ attends $p_2$ is realized with probability $\phi^b(\bar{b}^1)\psi(\omega_1)$. 

The interpretation of the model is therefore a random allocation over deterministic matchings. Each deterministic cutoff and signal realization leads to a deterministic matching, and the pair $(\pi, b)$ encodes the probability with which each deterministic matching is realized. In the rest of the paper, I refer to $(\pi, b)$ as a \textbf{matching}.

    \section{Main Result}
    In this section, I begin the analysis by introducing the market-clearing condition. Unlike the canonical setting, I define market clearing from an ex-ante perspective: given an ex-ante cutoff, the expected demand for each program equals its capacity. I then show that, for any signal profile, there exists an ex-ante cutoff that clears the market. In the second part of this section, I discuss how the setting of this paper relates to the existing literature. I then show that the corresponding standard results can also be established.

\subsection{Market-Clearing}
To establish the market-clearing, I first define the demand function of each program as follows:

\begin{definition}\label{def:demand}
    Given a signal profile $\pi$, the demand function $D^\pi_p: \mathbb{R}^n_{\ge 0} \to \mathbb{R}_{\ge 0}$ for each program $p$ is defined as follows:
    \begin{align*}
        D^\pi_p(b) = \sum_{\bar{b},k,i_k,\omega}\phi^b(\bar{b}) \psi(\omega) \pi_k(i_k|\omega)\ind{\sigma_k(C(s_k,\bar{b}),i_k) = p}  
    \end{align*}
\end{definition}

The demand is defined as a function of the ex-ante cutoff $b$. For each ex-ante cutoff $b$ and its associated distribution $\phi^b$, I aggregate the expected demand from all students, taking into account the probabilities of state and signal realizations. Because of preference uncertainty, demand is defined from an expected perspective. For each realized deterministic cutoff, state, and signal, the actual demand may fluctuate. Nevertheless, the \textbf{market-clearing} condition is also defined ex-ante, as introduced below:

\begin{definition}
    Fix a signal profile $\pi$, an ex-ante cutoff $b$ clears the market if
           \[
            D^\pi_p(b) \le M_p
           \]
          for all $p \in P$, with the equality holding if $b_p < t$. 
\end{definition}

Let $MC(\pi)$ denote the collection of ex-ante cutoffs that can clear the market under $\pi$. A matching $(\pi, b)$ is \textbf{feasible} under $\pi$ if $b \in MC(\pi)$.

Note that $t$ denotes the total number of students. The definition thus implies that each program must exactly fill its capacity, provided it does not admit all students. How I define the market-clearing condition captures the idea that, although programs cannot control fluctuations in demand across realized states and signals, they can derive an ex-ante cutoff such that the expected demand equals their capacity. In some realizations, too many students may accept offers, causing programs to exceed capacity; in others, too few students may enroll, leaving unfilled spots. These outcomes depend on the realized states of the world and signals.

Nevertheless, the ex-ante market-clearing concept provides a potential solution to problems that arise in matching markets with preference uncertainty. Before presenting the results, I first address the question of why—beyond preference uncertainty—I introduce an additional layer of uncertainty through probabilistic offers. In the following example, I show that when programs are restricted to using only deterministic cutoffs, the market may fail to clear, illustrating that introducing probabilistic offers is essential to achieving market clearing when preference uncertainty is present.

\begin{example}
 Let $P = \{p_1, p_2\}$ and $S = \{s_1, s_2\}$, and suppose both programs share the same priority ordering $s_1 \mathcal{R}_p s_2$, each with capacity one. Given the common prior $\psi$, the preference profiles of students and their corresponding states are as follows:

 \begin{table}[h!]
\centering
\begin{tabular}{|c|c|c|}
\hline
 &  $s_1$ & $s_2$ \\ \hline
$\omega_1$ & $p_1 \succ p_2$ & $p_1 \succ p_2$ \\ \hline
$\omega_2$ & $p_2 \succ p_1$ & $p_1 \succ p_2$ \\ \hline

\end{tabular}
\caption{Preference profiles of students with associated states}
\end{table}

Consider the signal profile $\pi$ that fully reveals the state for both students. Since each program should admit at least one student, the possible deterministic cutoffs are $(1,1)$, $(1,2)$, $(2,1)$, and $(2,2)$. Under $(1,1)$, the expected demand for each program is $(\psi(\omega_1), \psi(\omega_2)) \neq (1,1)$, which fails to clear the market. If $(1,2)$ or $(2,1)$ is chosen, the demand becomes $(\psi(\omega_1), \psi(\omega_2) + 1)$ or $(\psi(\omega_1) + 1, \psi(\omega_2))$, respectively, meaning that either $p_1$ or $p_2$ exceeds its capacity. Under $(2,2)$, the demand is $(\psi(\omega_1) + 1, \psi(\omega_2))$, so $p_1$ still exceeds its capacity. Hence, market clearing cannot be achieved through deterministic cutoffs.

\end{example}

Fortunately, once we introduce the random allocation for admissions, the market-clearing condition can always be achieved. The result is presented in the following proposition.

\begin{proposition}\label{prop:exist}
          
     For any signal profile $\pi$,  $MC(\pi)$, is nonempty. That is, there always exists an ex-ante cutoff that clears the market. Moreover, $MC(\pi)$ forms a complete lattice.
\end{proposition}

     The proof relies on the Knaster--Tarski theorem. We first observe that, once each student’s decision rule is given by maximizing expected utility, the induced demand function $D_p^\pi$ satisfies the gross substitutes condition. That is, the demand for program $p$ increases as $p$ admits more students while the cutoffs of other programs are fixed, and decreases as other programs admit more students while $p$'s cutoff is fixed. With the property of  gross substitutes, we can construct a monotone mapping that updates the cutoffs of programs with excess demand while holding others fixed. By the Knaster--Tarski theorem, this mapping admits a fixed point, which corresponds to the market-clearing cutoff.

      \subsection{Discussion}

      The main difficulty in matching markets with preference uncertainty lies in defining an appropriate notion of stability. Since students may have different preferences across states, the canonical definition of stability is no longer applicable. In this paper, I circumvent this difficulty by introducing expectations into the analysis, including expected demand and expected utility.

In the standard school choice problem, stability consists of two components. The first is the capacity constraint, which requires that the number of students that each school is matched with is less than or equal to its capacity. The second is the no-envy condition: if a student prefers some school to the one she is matched with, that school must be at full capacity, and the student must have lower priority than every student assigned to that school.

I formulate the capacity constraint using the market-clearing condition defined in this section, which guarantees that the expected number of students matched to each program is less than or equal to its capacity. As for the no-envy condition, it is automatically satisfied under the following assumptions. I assume that students are expected utility maximizers. Combined with the induced decision rule and the assumption that programs admit students sequentially according to their priority order, this ensures that there is no case in which a student $s$ would wish to be matched with some program $p$ with higher probability while another student $s'$ such that $s \,\mathcal{R}_p\, s'$ is matched with $p$ with positive probability.

In this paper, the concept of stability is formulated from an ex-ante perspective. Given that the signal profile is exogenously determined, it is treated as a fixed structure of uncertainty in the matching market. Accordingly, the definition of feasible matchings under each signal profile is analogous to the notion of stable matchings in the existing literature. Furthermore, the standard results in the matching literature can also be established analogously within this ex-ante framework. I begin by presenting the corresponding Rural Hospital Theorem in the following proposition.

      \begin{proposition}\label{prop:rural}
          Given a signal profile $\pi$, for all $b, \ b' \in MC(\pi)$, we have $D_p^\pi(b) = D_p^\pi(b')$ for all $p \in P$. Moreover, if $D_p^{\pi}(b)=D_p^{\pi}(b')<M_p$, then $p$ receives the same random allocation of students under $b$ and $b'$.
      \end{proposition}

      From the Rural Hospital Theorem, we know that each school is matched with the same number of students across all stable matchings. Moreover, any school that does not fill its capacity is always matched with the same set of students. In \cref{prop:rural}, each cutoff $b$ in the set $MC(\pi)$ corresponds to a feasible matching under the signal profile $\pi$. I claim that each program has the same expected demand across all feasible matchings. Although the set of students each program is matched with is not determined for each feasible matching, programs that do not fill their capacity always receive the same lottery over the set of students across all feasible matchings.

     Another standard result in the canonical setting is that the set of stable matchings forms a complete lattice under the Pareto order of students’ welfare, and a student-optimal stable matching exists. We can establish similar results from the perspective of expected utility. In our framework, each matching $(\pi, b)$ specifies a random allocation for every student. A student $s_k$ prefers one matching to another if the former yields a higher expected utility. Let $U_k(\pi, b)$ denote the expected utility of $s_k$ under matching $(\pi, b)$. Formally, 
\[
U_k(\pi, b) = \sum_{\bar{b}, i_k, \omega} \phi^b(\bar{b}) \psi(\omega) \pi_k(i_k \mid \omega) 
\, u_k\big(\sigma_k(C(s_k, \bar{b}), i_k), \omega\big).
\]



With the above notation in place, I am ready to present results regarding welfare analysis. We will continue to use this notation for the rest of the paper. First, I will present the connection between students' welfare and cutoffs: 
      \begin{proposition}\label{prop:lattice}
          Given a signal profile $\pi$, for all $b, \ b' \in \mathbb{R}^n_{\ge 0}$ such that $b \ge b' $, we have $U_k(\pi,b) \ge U_k(\pi,b')$ for all $1 \le k \le t$. That is, all students weakly prefer the higher cutoff under each signal profile.

      \end{proposition}

      Note that \cref{prop:lattice} is not restricted to the set of cutoffs that clear the market. 
This result is intuitive: while holding the information structure fixed, students achieve better outcomes when all programs admit weakly more students. 
If we restrict attention to the set of market-clearing cutoffs, the corresponding set of feasible matchings also forms a complete lattice under the Pareto order of students’ welfare. 
With additional welfare structures, we can establish an analogous result from the programs’ perspective. 
For completeness, these results are presented in the Appendix.

    \section{Information and Student Welfare}
    In this section, I discuss how the information structure affects student welfare. 
One might conjecture that a more informative signal profile should improve students’ welfare, 
as an extension of the Blackwell theorem. 
However, in our setting, the market-clearing cutoffs depend on the information structure, 
which implies that students may face different choice sets under different signal profiles. 
Consequently, while a student may receive more precise information under a new signal, 
her choice set may simultaneously shrink. 
In such cases, it is ambiguous whether the student actually benefits from the more informative signal. 
This setting departs from the canonical framework of choice under uncertainty, 
and therefore the Blackwell theorem does not directly apply. 
The relationship between informativeness and welfare is thus more subtle in this context.

\subsection{Preliminary}

Before presenting the results, I first introduce several definitions used in this section. 
To conduct the welfare analysis, I define the Pareto order of signal profiles as follows.

\begin{definition}
A signal profile $\pi$ is \textbf{Pareto dominated} by $\pi'$ if 
$U_k(\pi', \max MC(\pi')) \ge U_k(\pi, \max MC(\pi))$ for all $1 \le k \le t$, 
with at least one inequality strict. 
A signal profile $\pi$ is \textbf{Pareto efficient} if it is not dominated by any other signal profile.
\end{definition}

This definition is based on the expected utilities of students. 

Ex-post welfare comparisons are not considered in this paper. 
Intuitively, the Pareto order here is defined with respect to welfare before the realization of signals and cutoffs. 
Regarding signal comparisons, we know that the set of signals cannot be ranked by a complete order. I therefore adopt the standard Blackwell order for the analysis, introduced as follows.

\begin{definition}
A signal $\pi_k$ is more informative than another signal $\pi'_k$ 
if there exists a garbling function—that is, a conditional distribution $T(i'_k \mid i_k)$—such that
\[
    \pi'_k(i'_k \mid \omega) \;=\; \sum_{i_k} \pi_k(i_k \mid \omega)\, T(i'_k \mid i_k),
    \quad \forall\, i_k \in \text{Supp}(\pi_k), \ \omega \in \Omega.
\]
A signal profile $\pi$ is more informative than another signal profile $\pi'$, 
denoted by $\pi \succsim \pi'$, if $\pi_k$ is more informative than $\pi'_k$ for all $1 \le k \le t$.
\end{definition}

Note that when $\pi \succsim \pi'$, all students receive a more informative signal under $\pi$ 
than under $\pi'$, which is a strong condition. 
With these definitions in place, we can derive the following corollary.

\begin{cor}\label{cor}
A signal profile $\pi$ is not Pareto dominated by $\pi'$ 
if $\pi \succsim \pi'$ and $\max MC(\pi) \ge \max MC(\pi')$.
\end{cor}

By \cref{prop:lattice}, we know that all students weakly prefer the matching 
$(\pi', \max MC(\pi))$ to $(\pi', \max MC(\pi'))$. 
Moreover, students also weakly prefer the matching $(\pi, \max MC(\pi))$ 
to $(\pi', \max MC(\pi))$, because under $\pi$ they receive more informative signals 
while facing the same ex-ante cutoff as under $\pi'$. 
Hence, the Blackwell theorem applies, and \cref{cor} follows. 
However, the ex-ante cutoffs need not be monotone in informativeness, 
and the effect of the information structure on student welfare remains subtle.

\subsection{Information Can Hurt: An Example}

   Intuitively, even though a more informative signal profile may not Pareto improve students’ welfare, 
one might expect that it should not lead to a Pareto dominated outcome. 
Unfortunately, this conjecture is false: a more informative signal profile can, in fact, 
reduce the welfare of all students. Consider the following example.

Let $S = \{s_1, s_2, s_3, s_4\}$ and $P = \{p_1, p_2, p_3, p_4\}$, 
where each program has capacity one. 
The priorities of programs are constructed by:

\begin{table}[h!]
\centering
\begin{tabular}{|c|c|}
\hline
$p_1$ &  $s_3 \succ s_2 \succ s_1 \succ s_4$ \\ \hline
$p_2$ & $s_2 \succ s_1 \succ s_3 \succ s_4$ \\ \hline
$p_3$ & $s_1 \succ s_3 \succ s_2 \succ s_4$ \\ \hline
$p_4$ & $ s_4 \succ s_1 \succ s_2 \succ s_3$ \\
\hline

\end{tabular}
\caption{The priorities of programs}
\end{table}

The students’ preferences, together with the associated states and realized probabilities, 
are described in \cref{tab:ex1}.

\begin{table}[h!]
\centering
\begin{tabular}{|c|c|c|c|c|c|}
\hline
 & $\psi(\omega)$ & $s_1$ & $s_2$ & $s_3$ & $s_4$ \\ \hline
$\omega_1$ & 30\% & $p_1 \succ p_4 \succ p_2 \succ p_3$ & $p_2$ & $p_3 \succ p_1 \succ p_2 \succ p_4$ & $p_4 \succ p_3 \succ p_1 \succ p_2$ \\ \hline
$\omega_2$ & 40\% & $p_2 \succ p_3 \succ p_1 \succ p_4$ & $p_1$ & $p_3 \succ p_1 \succ p_2 \succ p_4$ & $p_4 \succ p_3 \succ p_1 \succ p_2$ \\ \hline
$\omega_3$ & 30\% & $p_3 \succ p_2 \succ p_1 \succ p_4$ & $p_1$ & $p_3 \succ p_1 \succ p_2 \succ p_4$ & $p_4 \succ p_3 \succ p_1 \succ p_2$ \\ \hline
\end{tabular}
\caption{Preferences of students with associated states and realized probabilities.}
\label{tab:ex1}
\end{table}

For $s_2$, I list only her top choice, as the rest of her ranking does not affect the example. 
The utility of each student $s_k$ is normalized as $u_k(p, \omega) = \bar{u}_i$, 
where $p$ is $s_k$’s $i$th choice under state $\omega$.

Consider a signal profile $\pi$ that gives all students the information partition 
$(\{\omega_1\}, \{\omega_2, \omega_3\})$. 
For each realization of this signal profile, each student has a distinct most-preferred program. 
Thus, we obtain $\max MC(\pi) = (4,4,4,4)$, and the associated choice sets and welfare levels are:

\begin{table}[h!]
\centering
\begin{tabular}{|c|c|c|}
\hline
 & Choice Set &  Expected Utility \\ \hline
$s_1$ & $P$ & $0.7 \bar{u}_1+0.3\bar{u}_2$ \\ \hline
$s_2$ & $P$ & $ \bar{u}_1$ \\ \hline
$s_3$ & $P$ & $ \bar{u}_1$ \\ \hline
$s_4$ & $P$ & $\bar{u}_1$ \\
\hline
\end{tabular}
\caption{Welfare of students with partial information.}
\end{table}

Under $\pi$, students $s_2$, $s_3$, and $s_4$ fully learn their preferences 
and always obtain their first-best outcome in each realized state. 
However, $s_1$ chooses $p_2$ when observing the partition $\{\omega_2, \omega_3\}$ 
and receives her first-best outcome with probability $\frac{4}{7}$. 
Accounting for state probabilities, $s_1$ attains her first-best outcome with probability $70\%$.

Now consider another signal profile $\pi'$ that fully discloses the state to all students. 
In this case, only $s_1$’s decision rule changes relative to $\pi$, 
since the other students already had full information under $\pi$. 
We derive $\max MC(\pi') = (2, 4, 1.7, 1)$, and the associated choice sets and welfare levels are:

\begin{table}[h!]
\centering
\begin{tabular}{|c|c|c|}
\hline
 & Choice Set &  Expected Utility \\ \hline
$s_1$ & $\{p_2,p_3\}$ & $0.7 \bar{u}_1+0.3\bar{u}_3$ \\ \hline
$s_2$ & $\{p_1, p_2\}$ & $ \bar{u}_1$ \\ \hline
$s_3$ & $0.7\{p_1, p_2, p_3\} + 0.3\{p_1, p_2\}$ & $ 0.7 \bar{u}_1 + 0.3 \bar{u}_2$ \\ \hline
$s_4$ & $\{p_2, p_4\}$ & $\bar{u}_1$ \\
\hline
\end{tabular}
\caption{Welfare of students with full information.}
\end{table} 

Here, $s_3$ receives choice set $\{p_1, p_2, p_3\}$ with probability $70\%$ 
and $\{p_1, p_2\}$ with probability $30\%$. 
All students now have strictly smaller choice sets under $\pi'$. 
While $s_2$ and $s_4$ maintain the same welfare level as under $\pi$, 
both $s_1$ and $s_3$ receive strictly lower welfare. 
Thus, even though $\pi' \succ \pi$, the signal profile $\pi'$ is Pareto dominated by $\pi$.

The intuition of how information hurts the welfare is as follows: 
when $s_1$ receives the more informative signal $\pi'$, 
she takes capacity from $p_3$, reducing $s_3$’s available options to $p_3$ 
since $s_3$ has lower priority under $p_3$. 
Therefore, $s_3$ can only take capacity from other programs and affect other students' choice set. This process
triggers a chain reaction that ultimately reduces $s_1$’s own choice set, and $s_1$ can only achieve a lower expected utility.

This example yields several insights. 
First, the ex-ante cutoff can fail to be monotone increasing in informativeness, 
and it can even be decreasing. 
Second, an outcome induced by a more informative signal profile 
can be Pareto dominated by one with less information, meaning that 
all students are weakly worse off (with at least one strictly worse off) 
despite the fact that all of them receive more information. 
Third, even full disclosure of information does not guarantee Pareto efficiency: 
Complete information may lead to worse welfare than partial information. 
Finally, a student (e.g., $s_1$ in this example) can harm her own welfare 
by learning more about her preferences.

It is important to emphasize that although each student’s decision 
does not affect others’ welfare once the ex-ante cutoffs are fixed, 
the cutoffs themselves are endogenously determined by the information structure. 
Hence, this setting differs fundamentally from the standard framework of choice under uncertainty. 
Information can reduce welfare, consistent with well-known results in game theory 
under imperfect information. 
However, I assume that students are naive utility maximizers, 
meaning that they cannot strategically commit to specific information acquisition processes. 
Both students and programs behave mechanically in this framework. 
Thus, the above result characterizes the outcome induced under these assumptions, 
without considering strategic behavior.

\subsection{Common Priority}

 Given that the effect of information structure on welfare is difficult to analyze in the general setting, I impose additional restrictions on the priority or preference profiles to derive a more positive result. A common simplification is to assume that all programs share the same priority over students; that is, $\mathcal{R}_p = \mathcal{R}_{p'}$ for all $p, p' \in P$. Under this assumption, we obtain the following result.

\begin{proposition}\label{prop:pareto}
    When all programs share the same priority, the signal profile that fully discloses the state of the world to all students is Pareto efficient.
\end{proposition}

Once the priority order is identical across all programs, the school choice problem reduces to a serial dictatorship under expected utility maximization. The student with the highest priority holds the choice set of the entire set of programs and can always choose her most preferred one, regardless of the realized state. There exists no alternative decision rule induced by any other signal that can strictly improve her utility. Hence, any other signal that yields the same utility for her must induce the same decision rule. The same reasoning applies recursively to the student with the second-highest priority, except that some program capacities are already filled by the first students, resulting in a smaller choice set. Repeating this argument sequentially according to the priority order establishes that no alternative signal profile can yield a strictly higher utility for any student without reducing others’ utilities compared to the fully revealing signal profile.

    \section{Concluding Remark}

    Uncertainty in preferences is a common feature across many types of matching markets, and recent studies have approached this issue through various methodologies. This paper focuses on a decentralized many-to-one matching market in which applicants learn about their preferences only after institutions have proposed a matching to them. When institutions are allowed to issue probabilistic offers, there exists an ex-ante cutoff that guarantees ex-ante market clearing, thereby achieving an ex-ante notion of stability. Since uncertainty is inevitable, targeting ex-post market clearing and stability is often unrealistic. My results, therefore, provide a new solution concept for such environments and offer a potential explanation for the widespread use of waitlist policies by institutions.

While I show that a market-clearing cutoff exists for all information structures, the mapping from information structures to market-clearing cutoffs remains unclear. Because these cutoffs do not vary monotonically with the informativeness of signals, the relationship between information structure and students’ welfare is subtle. An applicant may prefer to receive a noisier signal or even ignore certain information, since a more informative signal can strictly reduces her expected utility.

Although this paper does not derive definitive results regarding the welfare effects of information, this is an important topic that merits further study. One direction is to investigate whether specific structures of priorities or preferences can guarantee the monotonicity of market-clearing cutoffs or welfare with respect to informativeness. Another interesting avenue is to consider strategic information acquisition. In my current framework, applicants behave mechanically: they passively receive information and accept the offer that yields the highest expected utility given their posterior beliefs. However, in reality, applicants may be able to choose among different information structures or even commit to discarding information. Since these choices can influence the outcomes of all participants, studying the strategic selection of information structures and the resulting equilibrium behavior would be an intriguing extension.

\nocite{*}  
\printbibliography 

@article{ehlers2015matching,
  title={Matching markets under (in) complete information},
  author={Ehlers, Lars and Mass{\'o}, Jordi},
  journal={Journal of economic theory},
  volume={157},
  pages={295--314},
  year={2015},
  publisher={Elsevier}
}

@article{kloosterman2020school,
  title={School choice with asymmetric information: Priority design and the curse of acceptance},
  author={Kloosterman, Andrew and Troyan, Peter},
  journal={Theoretical Economics},
  volume={15},
  number={3},
  pages={1095--1133},
  year={2020},
  publisher={Wiley Online Library}
}

@article{liu2014stable,
  title={Stable matching with incomplete information},
  author={Liu, Qingmin and Mailath, George J and Postlewaite, Andrew and Samuelson, Larry},
  journal={Econometrica},
  volume={82},
  number={2},
  pages={541--587},
  year={2014},
  publisher={Wiley Online Library}
}

@article{liu2020stability,
  title={Stability and Bayesian consistency in two-sided markets},
  author={Liu, Qingmin},
  journal={American Economic Review},
  volume={110},
  number={8},
  pages={2625--2666},
  year={2020},
  publisher={American Economic Association 2014 Broadway, Suite 305, Nashville, TN 37203}
}

@article{chen2022information,
  title={Information acquisition and provision in school choice: a theoretical investigation},
  author={Chen, Yan and He, YingHua},
  journal={Economic Theory},
  volume={74},
  number={1},
  pages={293--327},
  year={2022},
  publisher={Springer}
}

@article{hastings2008information,
  title={Information, school choice, and academic achievement: Evidence from two experiments},
  author={Hastings, Justine S and Weinstein, Jeffrey M},
  journal={The Quarterly journal of economics},
  volume={123},
  number={4},
  pages={1373--1414},
  year={2008},
  publisher={MIT Press}
}

@article{arteaga2022smart,
  title={Smart matching platforms and heterogeneous beliefs in centralized school choice},
  author={Arteaga, Felipe and Kapor, Adam J and Neilson, Christopher A and Zimmerman, Seth D},
  journal={The Quarterly Journal of Economics},
  volume={137},
  number={3},
  pages={1791--1848},
  year={2022},
  publisher={Oxford University Press}
}

@article{fershtman2017pandora,
  title={Pandora's Auctions: Dynamic Matching with Unknown Preferences},
  author={Fershtman, Daniel and Pavan, Alessandro},
  journal={American Economic Review},
  volume={107},
  number={5},
  pages={186--190},
  year={2017},
  publisher={American Economic Association 2014 Broadway, Suite 305, Nashville, TN 37203}
}

@inproceedings{aziz2016stable,
  title={Stable matching with uncertain linear preferences},
  author={Aziz, Haris and Bir{\'o}, P{\'e}ter and Gaspers, Serge and De Haan, Ronald and Mattei, Nicholas and Rastegari, Baharak},
  booktitle={International Symposium on Algorithmic Game Theory},
  pages={195--206},
  year={2016},
  organization={Springer}
}

@article{dai2021learning,
  title={Learning strategies in decentralized matching markets under uncertain preferences},
  author={Dai, Xiaowu and Jordan, Michael I},
  journal={Journal of Machine Learning Research},
  volume={22},
  number={260},
  pages={1--50},
  year={2021}
}

@article{liu2015two,
  title={A Two-Sided Matching Decision Model Based on Uncertain Preference Sequences},
  author={Liu, Xiao and Ma, Huimin},
  journal={Mathematical Problems in Engineering},
  volume={2015},
  number={1},
  pages={241379},
  year={2015},
  publisher={Wiley Online Library}
}

@article{che2016decentralized,
  title={Decentralized college admissions},
  author={Che, Yeon-Koo and Koh, Youngwoo},
  journal={Journal of Political Economy},
  volume={124},
  number={5},
  pages={1295--1338},
  year={2016},
  publisher={University of Chicago Press Chicago, IL}
}

@article{bade2015serial,
  title={Serial dictatorship: The unique optimal allocation rule when information is endogenous},
  author={Bade, Sophie},
  journal={Theoretical Economics},
  volume={10},
  number={2},
  pages={385--410},
  year={2015},
  publisher={Wiley Online Library}
}

@article{bikhchandani2017stability,
  title={Stability with one-sided incomplete information},
  author={Bikhchandani, Sushil},
  journal={Journal of Economic Theory},
  volume={168},
  pages={372--399},
  year={2017},
  publisher={Elsevier}
}

@article{blair1988lattice,
  title={The lattice structure of the set of stable matchings with multiple partners},
  author={Blair, Charles},
  journal={Mathematics of operations research},
  volume={13},
  number={4},
  pages={619--628},
  year={1988},
  publisher={INFORMS}
}

@misc{avery2007new,
  title={The new market for federal judicial law clerks},
  author={Avery, Christopher and Jolls, Christine and Posner, Richard and Roth, Alvin E},
  year={2007},
  publisher={National Bureau of Economic Research Cambridge, Mass., USA}
}

@article{roth1985college,
  title={The college admissions problem is not equivalent to the marriage problem},
  author={Roth, Alvin E},
  journal={Journal of economic Theory},
  volume={36},
  number={2},
  pages={277--288},
  year={1985},
  publisher={Elsevier}
}

@article{abdulkadirouglu2003school,
  title={School choice: A mechanism design approach},
  author={Abdulkadiro{\u{g}}lu, Atila and S{\"o}nmez, Tayfun},
  journal={American economic review},
  volume={93},
  number={3},
  pages={729--747},
  year={2003},
  publisher={American Economic Association}
}

@article{abdulkadirouglu2009strategy,
  title={Strategy-proofness versus efficiency in matching with indifferences: Redesigning the NYC high school match},
  author={Abdulkadiro{\u{g}}lu, Atila and Pathak, Parag A and Roth, Alvin E},
  journal={American Economic Review},
  volume={99},
  number={5},
  pages={1954--1978},
  year={2009},
  publisher={American Economic Association}
}

@article{gale1962college,
  title={College admissions and the stability of marriage},
  author={Gale, David and Shapley, Lloyd S},
  journal={The American Mathematical Monthly},
  volume={69},
  number={1},
  pages={9--15},
  year={1962},
  publisher={Taylor \& Francis}
}

@article{kelso1982job,
  title={Job matching, coalition formation, and gross substitutes},
  author={Kelso Jr, Alexander S and Crawford, Vincent P},
  journal={Econometrica: Journal of the Econometric Society},
  pages={1483--1504},
  year={1982},
  publisher={JSTOR}
}

@article{roth1982economics,
  title={The economics of matching: Stability and incentives},
  author={Roth, Alvin E},
  journal={Mathematics of operations research},
  volume={7},
  number={4},
  pages={617--628},
  year={1982},
  publisher={INFORMS}
}

@article{grenet2022preference,
  title={Preference discovery in university admissions: The case for dynamic multioffer mechanisms},
  author={Grenet, Julien and He, YingHua and K{\"u}bler, Dorothea},
  journal={Journal of Political Economy},
  volume={130},
  number={6},
  pages={1427--1476},
  year={2022},
  publisher={The University of Chicago Press Chicago, IL}
}

@article{kapor2020heterogeneous,
  title={Heterogeneous beliefs and school choice mechanisms},
  author={Kapor, Adam J and Neilson, Christopher A and Zimmerman, Seth D},
  journal={American Economic Review},
  volume={110},
  number={5},
  pages={1274--1315},
  year={2020},
  publisher={American Economic Association 2014 Broadway, Suite 305, Nashville, TN 37203}
}

@article{chakraborty2010two,
  title={Two-sided matching with interdependent values},
  author={Chakraborty, Archishman and Citanna, Alessandro and Ostrovsky, Michael},
  journal={Journal of Economic Theory},
  volume={145},
  number={1},
  pages={85--105},
  year={2010},
  publisher={Elsevier}
}

@article{chen2020learning,
  title={Learning by matching},
  author={Chen, Yi-Chun and Hu, Gaoji},
  journal={Theoretical Economics},
  volume={15},
  number={1},
  pages={29--56},
  year={2020},
  publisher={Wiley Online Library}
}

@article{chen2021information,
  title={Information acquisition and provision in school choice: an experimental study},
  author={Chen, Yan and He, Yinghua},
  journal={Journal of Economic Theory},
  volume={197},
  pages={105345},
  year={2021},
  publisher={Elsevier}
}

@article{immorlica2020information,
  title={Information acquisition in matching markets: The role of price discovery},
  author={Immorlica, Nicole and Leshno, Jacob and Lo, Irene and Lucier, Brendan},
  journal={Available at SSRN 3705049},
  year={2020}
}

@article{escobar2024search,
  title={Search and Information in Centralized School Choice Systems},
  author={Escobar, Juan F and Montes, Alfonso},
  year={2024}
}

@article{kamenica2011bayesian,
  title={Bayesian persuasion},
  author={Kamenica, Emir and Gentzkow, Matthew},
  journal={American Economic Review},
  volume={101},
  number={6},
  pages={2590--2615},
  year={2011},
  publisher={American Economic Association}
}

@article{gentzkow2016rothschild,
  title={A Rothschild-Stiglitz approach to Bayesian persuasion},
  author={Gentzkow, Matthew and Kamenica, Emir},
  journal={American Economic Review},
  volume={106},
  number={5},
  pages={597--601},
  year={2016},
  publisher={American Economic Association 2014 Broadway, Suite 305, Nashville, TN 37203}
}

@article{azevedo2016supply,
  title={A supply and demand framework for two-sided matching markets},
  author={Azevedo, Eduardo M and Leshno, Jacob D},
  journal={Journal of Political Economy},
  volume={124},
  number={5},
  pages={1235--1268},
  year={2016},
  publisher={University of Chicago Press Chicago, IL}
}

@article{bergemann2016bayes,
  title={Bayes correlated equilibrium and the comparison of information structures in games},
  author={Bergemann, Dirk and Morris, Stephen},
  journal={Theoretical Economics},
  volume={11},
  number={2},
  pages={487--522},
  year={2016},
  publisher={Wiley Online Library}
}

\appendix

\section{More General Form of Decision Rule}

I refer to the decision rule defined in the main text as the naive decision rule, where its output must be a singleton from the given choice set. However, a student is also allowed to randomize her decision over the choice set. Moreover, the naive decision rule depends solely on the signal realization. In general, a decision rule can be viewed as a recommendation to each student, meaning that it could depend on both the state of the world and the corresponding signal realization. In this appendix, I provide a more general definition of a decision rule, as follows.

A decision rule with respect to a choice set $C$ for student $s_k$ with signal $\pi_k$ is a mapping $
\sigma_k(C): I_k \times \Omega \to \Delta(C)$, where $\sigma_k(C)(p| i_k, \omega)$ denotes the probability that $s_k$ chooses program $p \in C$, given state $\omega$ and signal realization $i_k$. We can think of $\sigma_k(C)$ as a ``black box'' that outputs a recommendation to $s_k$ under different states and signal realizations. A decision profile is denoted by $
\sigma = (\sigma_k(C))_{C \in 2^P, \ 1 \le k \le t}$. Now, I introduce two essential properties of decision rule for the paper.

\paragraph{Obedience.}
Given a signal $\pi_k$, a decision rule $\sigma_k(C)$ is obedient under $\pi_k$ \parencite{bergemann2016bayes} if, for all $i_k \in \text{Supp}(\pi_k)$,
\begin{align*}
& \sum_{\omega \in \Omega} \psi(\omega)\pi_k(i_k | \omega)\sigma_k(C)(p | i_k,\omega)u_k(p,\omega) \\
\ge\ & \sum_{\omega \in \Omega} \psi(\omega)\pi_k(i_k | \omega)\sigma_k(C)(p' | i_k,\omega)u_k(p',\omega)
\quad \forall \, p, p' \in C.
\end{align*}
This inequality states that, if $s_k$ is recommended to take program $p$ upon observing signal $i_k$, she cannot be better off by choosing any other program $p'$. A decision profile $\sigma$ satisfies the obedience condition under $\pi$ if $\sigma_k(C)$ is obedient under $\pi_k$ for all $C \in 2^P$ and $1 \le k \le t$.

\paragraph{Gross Substitutes.}
Given a signal $\pi_k$, a decision rule $\sigma_k$ satisfies gross substitutes if, for all $C, C' \in 2^P$ such that $C \subseteq C'$,
\[
\sigma_k(C)(p | i_k,\omega) \ge \sigma_k(C')(p | i_k,\omega)
\quad \forall \, p \in C, \ i_k \in \text{Supp}(\pi_k), \ \omega \in \Omega.
\]
This property guarantees that, holding the information structure fixed, expanding the choice set of $s_k$ should not increase the probability assigned to any program already in the smaller choice set. Intuitively, once a student has more options, her demand for the alternatives initially available should not increase. A decision profile $\sigma$ satisfies gross substitutes under $\pi$ if $\sigma_k$ satisfies this property under $\pi_k$ for all $1 \le k \le t$.

Let $OG(\pi)$ denote the set of decision profiles that satisfy both the obedience condition and gross substitutes under the signal profile $\pi$. Note that the signal profile $\pi$ is exogenously given—we can interpret it as disclosing information from a subjective perspective, beyond anyone’s control (e.g., whether a student enjoys interacting with faculty during the visit days). The decision rule, on the other hand, can be viewed as disclosing information from an objective perspective. We may imagine a third party with full information who decides what to reveal to students—for example, historical placement outcomes of graduate students. Hence, programs or this third party can design the decision rule for each student to disclose additional information beyond the signal $\pi$, while the obedience constraint ensures incentive compatibility—specifically, the decision rule will not recommend any action that yields a lower expected utility than naively choosing the program that provides the highest expected utility under the posterior induced by the signal realization.

The naive decision rule clearly satisfies both gross substitutes and the obedience condition. It is the simplest case that the decision rule provides no additional information beyond the signal $\pi$. For simplicity, I present my main results using this naive decision rule in the main text. However, given a signal profile $\pi$, my results extend to any decision rule in $OG(\pi)$, and the proofs are written under this general formulation.

\section{Omitted Proof}

In the following proof, I formulate the idea of random allocation by assuming that students are divisible into a continuum of infinitesimal parts. Formally, each student $s_k$ is represented as the set
\[
s_k = \{(k,e): e \in [0,1]\},
\]
which has measure $1$. I call each element $(k,e)$ a student point. Each student point $(k,e)$ shares the same strict preference $\succ_k$ over the set of programs $P$ as $s_k$.

Similarly, I rewrite the structure of ranks, cutoffs, and choice sets under this alternative notion. Each program $p$ assigns a rank $r_p(k,e)$ to each student point $(k,e)$, where 
\[
r_p(k,e) = |\{s_{k'} \in S : s_{k'} \succ_p s_k\}| + e, \quad \forall p \in P.
\]
An ex-ante cutoff $b = (b_1, \ldots, b_n) \in [0,t]^n$ specifies the total measure of student points admitted by each program. A student point $(k,e)$ is admitted to program $p$ under the cutoff $b$ if \(r_p(k,e) \leq b_p\). Define $
C: \mathbb{N} \times [0,1] \times [0,t]^n \to 2^P
$ as the choice set function, where 
\[
C(k,e,b) = \{p \in P : r_p(k,e) \le b_p\}
\]
denotes the choice set of the student point $(k,e)$ under the cutoff $b$. We are now prepared to define market clearing under this alternative formulation.

\begin{definition}
    Fix a pair of signal profile and decision profile $(\pi,\sigma)$. The demand function $D_p^{(\pi,\sigma)}: \mathbb{R}^n_{\ge 0} \to \mathbb{R}_{\ge 0}$ for each program $p$ is defined as
    \begin{align*}
        D_p^{(\pi,\sigma)}(b) 
        = \sum_{k,i_k,\omega} \int_0^1 
        \psi(\omega) \pi_k(i_k|\omega) 
        \sigma_k(C(k,e,b))(p|i_k,\omega) \, de.
    \end{align*}
\end{definition}

Given an ex-ante cutoff $b$, the choice set of each student point $(k,e)$ is determined, and the corresponding probability of selecting each program is given by $\sigma_k(C(k,e,b))$. The demand function of program $p$ aggregates the measure of student points that receive offers from $p$ and their corresponding probabilities of selecting $p$ from their choice sets. This notion is equivalent to \cref{def:demand} in the main text.  

Since programs are only allowed to randomize between two deterministic cutoffs that differ by one, representing students as a continuum of points is equivalent to random allocation. Given a choice set $C$, for each student $s_k$, the measure of student points $(k,e)$ allocated with $C$ can be interpreted as the probability that $s_k$ receives the choice set $C$. We can now define the market clearing condition under this formulation.

\begin{definition}
    Fix a pair of signal profile and decision profile $(\pi,\sigma)$.  
    An ex-ante cutoff $b$ clears the market under $\sigma$ and $\pi$ if
    \[
        D_p^{(\pi,\sigma)}(b) \le M_p
    \]
    for all $p \in P$, with equality whenever $b_p < t$.  
    Let $MC(\pi, \sigma)$ denote the collection of ex-ante cutoffs that clear the market under $\pi$ and $\sigma$.
\end{definition}

Now, we are ready to present the proof. The existence of market clearing and the rural hospital theorem follow arguments similar to those in \textcite{azevedo2016supply}. For the proof of \cref{prop:pareto}, I do not extend the result to the general form of the decision rule.

\subsection{Proof of \cref{prop:exist}}

\begin{proposition}
          For each pair of $(\pi,\sigma)$, such that $\sigma \in OG(\pi)$, $MC(\pi,\sigma)$ is not empty. Moreover, $MC(\pi,\sigma)$ forms a complete lattice.
\begin{proof}

For simplicity, we drop the superscript in the followings. We first argue that the demand function for each program $p \in P$, $D_p(b)$ satisfies the following property: $D_p(b_p,b_{-p}) \ge D_p(b_p',b_{-p})$ for all $b_p \ge b_p'$, and $ D_p(b_p,b_{-p}) \ge D_p(b_p,b'_{-p})$ for all $b_{-p} \le b_{-p}'$. 

To show first inequality, let $b  = (b_p,b_{-p})$ and $b'  = (b_p',b_{-p})$. For each student point $(k,e)$, it is either the case that $C(k,e,b) = C(k,e,b')$ or $C(k,e,b)\cup p = C(k,e,b')$. For the first type of student pints, we have $\sigma_k(C(k,e,b)) = \sigma_k(C(k,e,b'))$. As for the other type of student points, we have $\sigma_k(C(k,e,b))(p|i_k,\omega) \ge \sigma_k(C(k,e,b'))(p|i_k,\omega) = 0$ for all $i_k \in Supp(\pi_k)$ and $\omega \in \Omega$. Hence, we have $D_p(b_p,b_{-p}) \ge D_p(b_p',b_{-p})$ for all $b_p \ge b_p'$.
           
For the other inequality, let $b  = (b_p,b_{-p})$ and $b'  = (b_p,b'_{-p})$, where $b_{-p} \le b_{-p}'$. For each student $(k,e)$, it must be the case that $C(k,e,b) \subseteq C(k,e,b')$. If $p \in C(k,e,b) \subseteq C(k,e,b')$, by the assumption of the gross substitute, we have $\sigma_k(C(k,e,b))(p|i_k,\omega) \le \sigma_k(C(k,e,b'))(p|i_k,\omega)$ for all $i_k \in Supp(\pi_k)$ and $\omega \in \Omega$. If $p \not\in C(k,e,b)$ then $ p \not\in C(k,e,b')$, since the cutoff of $p$ are the same between $b$ and $b'$. Under this case we have $\sigma_k(C(k,e,b))(p|i_k,\omega) = \sigma_k(C(k,e,b'))(p|i_k,\omega) = 0$ for all $i_k \in Supp(\pi_k)$ and $\omega \in \Omega$. Hence, we have $D_p(b_p,b_{-p}) \le D_p(b_p',b_{-p})$ for all $b_p \ge b_p'$. 

Now, we define the operator $T = (T_1,...,T_n)$ as follows:
           \begin{align*}
               I_p(b_{-p}) &= \{b_p' \in [0,t]|D_p(b_p',b_{-p}) \le M_p \ \text{and the equality holds if}  \ b_p'<t\}\\
               T_p(b) &= \max I_p(b_{-p})
           \end{align*}
           We now prove that $T$ is well-defined and monotone. Note that, because $D_p(0, b_{-p}) = 0$ and $D_p$ is continuous, then either there exists $b_p \in [0,t]$ such that $D_p(b_p, b_{-p}) = M_p$ or $t \in I_p(b_{-p})$. In either case, we have that $I_p(b_{-p})$ is nonempty. Note that, by monotonicity and continuity of demand, $I_p(b_{-p})$ is a compact interval.

           To prove that $T$ is monotone, consider $b \le b'$, $t_p = T_p(b)$, and $t_p' = T_p(b')$. Assume that $t_p > t_p'$, and $t_p' < t$. By definition of $I_p$, we have
           \begin{align*}
               M_p = D_p(t_p',b'_{-p}) \le D_p(t_p,b'_{-p}) \le D_p(t_p,b_{-p}) \le M_p \\
               M_p = D_p(t_p',b'_{-p}) \le D_p(t_p',b_{-p}) \le D_p(t_p,b_{-p}) \le M_p
           \end{align*}
           We must have $M_p =  D_p(t_p,b'_{-p}) = D_p(t_p',b_{-p})$. This implies that $t_p \in I_p(b_{-p}')$. However, $t_p' < t_p$, and this contradicts to $T_p(b') = t_p'$. We have shown that $T$ is monotone. By Knaster–Tarski theorem, the fixed point of $T$ exists and achieves the market clearing. Moreover, the set of fixed points forms a complete lattice.

           \end{proof}
           \end{proposition}

\subsection{Proof of \cref{prop:rural}}

\begin{proposition}\label{prop:b.2}
          For each pair of $(\pi,\sigma)$, such that $\sigma \in OG(\pi)$, for any $b, \ b' \in MC(\pi,\sigma)$, we have $D_p^{(\pi,\sigma)}(b) = D_p^{(\pi,\sigma)}(b')$ for all $p_j \in P$. Moreover, if $D_p^{(\pi,\sigma)}(b)=D_p^{(\pi,\sigma)}(b')<M_p$, then $p$ receives the same set of student points under $b$ and $b'$.
\begin{proof}          

Consider $b, \ b' \in MC(\pi,\sigma)$, and let $b^- = b \wedge b'$. Choose $p \in P$, and assume without loss of generality that $b_p \ge b_p'$. We know that $D_p(b^-) \ge D_p(b')$, as $b^{-}_p = b_p'$ and $b^{-}_{-p} \le b_{-p}'$. If $b_p'<t$, then $D_p(b^-) = M_p \ge D_p(b)$. If $b_p' = t = b_p$, then $D_p(b^{-}) \ge D_p(b)$. In both cases, we can establish that
              \begin{align*}
                  D_p(b^-) \ge \max \{D_p(b), D_p(b') \}
              \end{align*}
              , and this result holds for all $p \in P$. 
              
Note that the measure of matched student and unmatched student should always sum up to $t$. Hence, the measure of the unmatched student under an ex-ante cutoff $b$ is $t - \sum_{p \in P} D_p(b)$. Since $b^-$ admits less student points than $b$ and $b'$, the measure of unmatched student points should be larger under $b^-$. We should have $t - \sum_{p \in P} D_p(b^-) \ge t - \sum_{p \in P} D_p(b)$ and $t - \sum_{p \in P} D_p(b^-) \ge t - \sum_{p \in P} D_p(b')$, which imply that $\sum_{p \in P} D_p(b^-) \le  \sum_{p \in P} D_p(b)$ and $\sum_{p \in P} D_p(b^-) \le \sum_{p \in P} D_p(b')$. Therefore, we must have $D_p(b^-) = D_p(b) = D_p(b')$ for all $p_j \in P$, or one of the above inequalities cannot hold.

For the second part of the proposition, take a program $p$ such that $D_p(b) = D_p(b')<M_p$. Let $b^- = b \wedge b'$. By the definition of market clearing, we know that $b_p = b_p' = b_p^- = t$. Combine this observation and $b^- = b \wedge b'$. We should have $p \in C(k,e,b^-) \subseteq C(k,e,b)$ and $p \in C(k,e,b^-) \subseteq C(k,e,b)$, for all student points $(k,e)$.
              
I first show that the demand of $p$ under each state and its signal realization is the same between $b$ and $b^-$. By the assumption that $\sigma \in OG(\pi)$, we know that, for all $1 \le k \le t$, $ \sigma_k(C(k,e,b))(p|i_k,\omega) \le \sigma_k(C(k,e,b^-))(p|i_k,\omega) \ \forall  \ i \in Supp(\pi_k), \ \omega \in \Omega$, since  $p \in C(k,e,b^-) \subseteq C(k,e,b)$. The measure of student points such that $C(k,e,b^-) \subset C(k,e,b)$ is positive, or there is no difference between $b$ and $b^-$. Therefore, if there exists a signal $i$ or a state $\omega$ such that the equality does not hold, we will have $D_p(b)<D_p(b^-)$, which contradicts to the first part of the proposition. Hence, we must have that, for all $(k,e)$, $ \sigma_k(C(k,e,b))(p|i_k,\omega) = \sigma_k(C(k,e,b^-))(p|i_k,\omega) \ \forall  \ i \in Supp(\pi_k), \ \omega \in \Omega$. We can also derive the result that for all $(k,e)$, $ \sigma_k(C(k,e,b'))(p|i_k,\omega) = \sigma_k(C(k,e,b^-))(p|i_k,\omega) \ \forall  \ i \in Supp(\pi_k), \ \omega \in \Omega$ by the same argument. We can conclude that each $(k,e)$ will take the program $p$ with the same probability under each state and its signal realization within $b$, $b'$, and $b^-$. 
              
              \end{proof}
              \end{proposition}

\subsection{Proof of \cref{prop:lattice}}

For the welfare analysis, we have to rewrite the expected utility of student and programs in the notion of the general form of decision rule. Fix a pair of $(\pi,\sigma)$, the expected utility of $s_k$ is a mapping from the set of ex-ante cutoff to a real number, where 

\begin{align*}
          U_k^{(\pi,\sigma)}(b) = \sum_{i_k,\omega,p} \int_0^1 \psi(\omega) \pi_k(i_k|\omega)\sigma_k(C(k,e,b))(p|i_k,\omega) u_k(p,\omega) de 
      \end{align*}

For each program $p$, I first define $u_p:S \to \mathbb{R}$ as its utility function, where $u_p(s_k)$ represents the utility of $p$ while receiving student $s_k$. We hold the assumption that $u_p(s_k) > u_p(s_{k'})$ if and only if $s_k \  \mathcal{R}_p \ s_{k'}$. Fix a pair of $(\pi,\sigma)$, the expected utility of $p$ is also a mapping from the set of ex-ante cutoff to a real number, defined as follows:
      \begin{align*}
          U_p^{(\pi,\sigma)}(b) = \sum_{k,i_k,\omega} \int_0^1 \psi(\omega) \pi_k(i_k|\omega)\sigma_k(C(k,e,b))(p|i,\omega) u_p(s_k) de
      \end{align*}
Now, I can present the result regarding the lattice structure of the set of ex-ante cutoffs with respect to the welfare.

\begin{proposition}
          Fix a pair of $(\pi,\sigma)$, such that $\sigma \in OG(\pi)$, for all $b, \ b' \in \mathbb{R}^n_{\ge 0}$ such that $b \le b' $, we have
          \begin{itemize}
              \item $U_k^{(\pi,\sigma)}(b) \le U_k^{(\pi,\sigma)}(b')$, for all $1 \le k \le t$
              \item $U_p^{(\pi,\sigma)}(b) \ge U_k^{(\pi,\sigma)}(b')$, for all $p \in P$
          \end{itemize}

          \begin{proof}
              We start by the first part of the proposition. For each $(k,e)$, we must have $C(k,e,b) \subseteq C(k,e,b')$. For simplicity, let $C(k,e,b) = C$ and $ C(k,e,b') = C'$. By the assumption of gross substitute, for all $p\in C$, we know that $\sigma_k(C')(p|i_k,\omega) \le \sigma_k(C)(p|i_k,\omega)$, for all $i_k \in Supp(\pi_k)$, $\omega \in \Omega$, and $1 \le k \le t$. Let $\sigma_k(C')(p|i_k,\omega) = \sigma_k(C)(p|i_k,\omega) - \epsilon_p(i_k,\omega)$, where $\epsilon_p (i_k,\omega) \ge 0$, for all $p \in C$. For each $(k,e)$ we have,
              \begin{align*}
                   &\sum_{\omega,i_k,p} \psi(\omega) \pi_k(i_k|\omega) \sigma_k(C')(p|i_k,\omega) u_k(p,\omega) \\
                = & \sum_{p \in C'\setminus C} \sum_{\omega,i_k} \psi(\omega) \pi_k(i_k|\omega) \sigma_k(C')(p|i_k,\omega) u_k(p,\omega) + \sum_{p \in C} \sum_{\omega,i_k} \psi(\omega) \pi_k(i_k|\omega) \sigma_k(C')(p|i_k,\omega) u_k(p,\omega)\\
                 \ge & \sum_{p \in C'\setminus C} \max_{p' \in C} \sum_{\omega,i_k} \psi(\omega) \pi_k(i_k|\omega) \sigma_k(C')(p|i_k,\omega) u_k(p',\omega) + \sum_{p \in C} \sum_{\omega,i_k} \psi(\omega) \pi_k(i_k|\omega) \sigma_k(C')(p|i_k,\omega) u_k(p,\omega)\\
                  = & \sum_{p \in C'\setminus C} \max_{p' \in C} \sum_{\omega,i_k} \psi(\omega) \pi_k(i_k|\omega) \sigma_k(C')(p|i_k,\omega) u_k(p',\omega) + \sum_{p \in C} \sum_{\omega,i_k} \psi(\omega) \pi_k(i_k|\omega) [\sigma_k(C)(p|i_k,\omega) - \epsilon_p(i_k,\omega)] u_k(p,\omega) \\
                  \ge & \sum_{p \in C}  \sum_{\omega,i_k} \psi(\omega) \pi_k(i_k|\omega) \epsilon_p(i_k,\omega) u_k(p',\omega) + \sum_{p \in C} \sum_{\omega,i_k} \psi(\omega) \pi_k(i_k|\omega) [\sigma_k(C)(p|i_k,\omega) - \epsilon_p(i_k,\omega)] u_k(p,\omega) \\
                  = & \sum_{\omega,i_k,p} \psi(\omega) \pi_k(i_k|\omega) \sigma_k(C)(p|i_k,\omega) u_k(p,\omega)
              \end{align*}
              The first inequality comes from the obedience condition. When $(k,e)$ is recommend to take some program $p \in C'\setminus C$, it cannot be better off by taking another program $p' \in C$. And the second inequality comes from the fact that choosing the $\epsilon_p(i_k,\omega)$ is simply a feasible alternative to the maximization term in the previous line. This conclusion  holds for all $(k,e)$. By aggregating all states and signal realization,  we can conclude that $U_k^{(\pi,\sigma)}(b) \le U_k^{(\pi,\sigma)}(b')$, for all $1 \le k \le t$.

              Now, I show the second part of the proposition. Fix a program $p$, for each $(k,e)$, there are three cases: $p \in C$ and $p \in C'$, $p \not\in C$ but $p \in C'$, or $p \not\in C$ and $p \not\in C'$. For the first kind of student points $(k,e)$, we know that $\sigma_k(C)(p|i_k,\omega) \le \sigma_k(C')(p|i_k,\omega)$ by the gross substitute assumption. For the second kind of student, have  $\sigma_k(C')(p|i,\omega) \ge \sigma_k(C)(p|i,\omega)=0$, since $p \in C$. As for the last kind of student points $(k,e)$, we have $\sigma_k(C')(p|i,\omega) = \sigma_k(C)(p|i,\omega)=0$, which does not affect the welfare. We conclude that, while moving from the cutoff $b$ to $b'$, the decrease demand of $p$ all come from the first kind of student points and the increase demand of $p$ all come from the second kind of student points.

              By \cref{prop:b.2}, we know that $D_p^{(\pi,\sigma)}(b) = D_p^{(\pi,\sigma)}(b')$. The measure of increment of student points and decrement of student points should be equal. We also know that for any pair of student point $(k,e)$ and $(k'.e')$, such that $(k,e)$ is the first kind, and $(k',e')$ is the second kind, we must have $u_p(s_k) \ge u_p(s_k')$. This is obvious since the second kind of student points are not admitted by $p$ under the lower cutoff $b$, which implies $r_p(k,e) \ge r_p(k',e')$.
              
              Moving from $b$ to $b'$, we replace a positive measure of student points (first kind) with the same measure of student points with lower priority. We can conclude that $U_p^{(\pi,\sigma)}(b) \ge U_k^{(\pi,\sigma)}(b')$.
          \end{proof}

      \end{proposition}

              \subsection{Proof of \cref{prop:pareto}}
              Let $\pi$ be the signal profile that fully discloses states to all students and assume that there exists a signal profile $\pi'$, such that $\pi$ is Pareto dominated by $\pi'$. 
              
              Without loss of generality, we assume that all programs share the same priority such that $s_1 \mathcal{R}s_2\mathcal{R}...\mathcal{R}s_t$. Let $A^k(e,\omega) \in \mathbb{R}^n$ denote the allocation of programs for the student point $(k,e)$ under the state $\omega$. For example, let $P= \{p_1,p_2\}$, $A^k(e,\omega) = (0.3,0.7)$ represents that $(k,e)$ will be allocated to $p_1$ and $p_2$ with probability $0.3$ and $0.7$ under $\omega$, respectively. Under the allocation $A^k$, we can write the expected utility of student $s_k$ as follows:
         \begin{align*}
             \sum_{\omega\in\Omega} \psi(\omega) \sum_{p\in P} \int_0^1 A_p^k(e,\omega) \cdot u_k(p,\omega) \ de
         \end{align*}
         We start by analyzing the welfare of $s_1$.  Under the full disclosure signal profile $\pi$, each student point will choose the program that it prefers the most under each state. Hence, the decision rule $\sigma_1$ associated with $\pi_1$ will solve the following welfare maximizing problem. 
         \begin{align*}
              U_1(\pi) = & \max_{A^1}\sum_{\omega\in\Omega} \psi(\omega) \sum_{p\in P} \int_0^1 A_p^1(e,\omega) \cdot u_1(p,\omega) \ de\\
              s.t.& \ \sum_pA_p^1(e,\omega) \le 1 \ \forall \ e \in [0,1], \ \omega \in \Omega\\
              & \sum_{\omega\in\Omega} \psi(\omega) \int_0^1 A_p^1(e,\omega) \ de^1 \le M_p \ \forall \ p\in P 
         \end{align*}

         Note that the set of the allocation $A_1$ that solves the maximization must be a singleton, since we assume the strict preference over programs in each state. By assumption, we should have $U_1(\pi') \ge U_1(\pi)$. The equality must hold, since $U_1(\pi)$ is the maximum value under the current environment. Under $\pi'$, we must assign the same allocation $A^1$ to $s_1$, or we will have $U(\pi') < U(\pi)$ and violates the assumption.

         Similarly, for each $s_k$ the allocation under the full disclosure will solve the following problem:
         \begin{align*}
              U_k(\pi) = & \max_{A^k}\sum_{\omega\in\Omega} \psi(\omega) \sum_{p\in P} \int_0^1 A_p^k(e,\omega) \cdot u_k(p,\omega) \ de\\
              s.t.& \ \sum_pA_p^k(e,\omega) \le 1 \ \forall \ e \in [0,1], \ \omega \in \Omega\\
              & \sum_{i=1}^{k-1}A^i_p +  \sum_{\omega\in\Omega} \psi(\omega) \int_0^1 A_p^k(e,\omega) \ de \le M_p \ \forall \ p\in P 
         \end{align*}

         , where
         \begin{align*}
             A^i_p = \sum_{\omega\in\Omega} \psi(\omega) \int_0^1 A_p^i(e,\omega) \cdot u_i(p,\omega) \ de
         \end{align*}

         represents the aggregate  demand of the program $p$ from $s_i$. We can interpret that $A^k$ is the best allocation that $s_k$ can have given that $A^i$ is fixed for all $i < k$. We have argued that we must assign the same allocation $A^1$ to $s_1$ under $\pi'$. This argument will holds for $s_2$, $s_3$,...,$s_t$. Since once we assign different allocation for $s_k$, we will have $U_k(\pi')<U_k(\pi)$. Finally, since $U_k(\pi) = U_k(\pi')$ for all $1 \le k \le t$. We conclude that $\pi$ is not Pareto dominated by $\pi'$. $\pi$ is Pareto efficient.

\end{document}